# Development of a plasma panel radiation detector: recent progress and key issues

Y. Silver, R. Ball, J. R. Beene, Y. Benhammou, M. Ben-Moshe, J. W. Chapman, T. Dai, E. Etzion, C. Ferretti, N. Guttman, P. S. Friedman, D. S. Levin, S. Ritt, R. L. Varner, C. Weaverdyck, B. Zhou

*Abstract*—A radiation detector based on plasma display panel technology, which is the principal component of plasma television displays is presented. Plasma Panel Sensor (PPS) technology is a variant of micropattern gas radiation detectors. The PPS is conceived as an array of sealed plasma discharge gas cells which can be used for fast response ($O$(5ns) per pixel), high spatial resolution detection (pixel pitch can be less than 100 micrometer) of ionizing and minimum ionizing particles. The PPS is assembled from non-reactive, intrinsically radiation-hard materials: glass substrates, metal electrodes and inert gas mixtures. We report on the PPS development program, including simulations and design and the first laboratory studies which demonstrate the usage of plasma display panels in measurements of cosmic ray muons, as well as the expansion of experimental results on the detection of betas from radioactive sources.

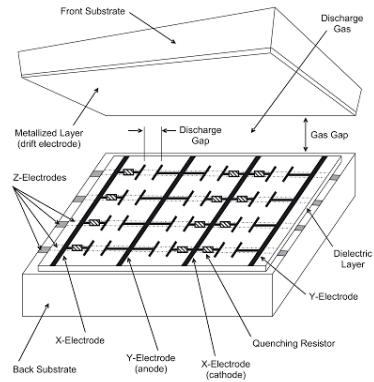

Fig. 1. Concept drawing (not to scale) of PPS cell electrode configuration. This is a 4-electrode structure. The X and Y lines define the cell discharge gap with embedded resistors actually under the X-electrodes (surface-discharge panel).

## I. Introduction

WE are investigating a new radiation detector technology based on plasma display panels (PDP), that are incorporated in large area, plasma television displays. The design and production of PDPs is supported by an extensive and experienced industrial base with four decades of development. The Plasma Panel Sensor (PPS) is a novel variant of the micropattern radiation detector [1][2][3][4], and should exploit the industrial and technology base of the plasma display panels. A PDP comprises millions of cells (3 cells form one pixel in a color PDP) per square meter, each of which, when provided with a signal pulse, can initiate and sustain a plasma discharge. This plasma discharge translates into the visible light emitted from the PDP (mostly due to phosphor material covering the inner side of each cell).

The light from each pixel in a PDP is emitted from a plasma created by an electric discharge. Discharge dimensions are in the 100 micrometer range typically at a pressure of 500 Torr, and the applied voltage between the electrodes of a few

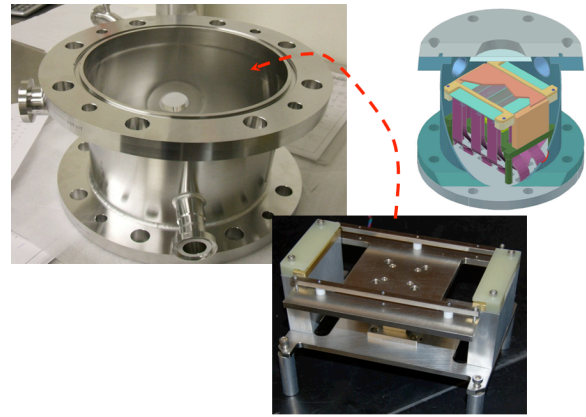

Fig. 2. Cut away view alongside with a picture of the fully configurable PPS test vessel and the motorized platform.

This work was supported in part by the U.S. Department of Energy under Grant No's: DE-FG02-07ER84749, DE-SC0006204, and DE-SC0006219. This work was also supported in part by the Office of Nuclear Physics, U. S. Department of Energy, and by the BSF under Grant No. 2008123.

Y. Silver, Y. Benhammou, M. Ben-Moshe, E. Etzion and N. Guttman are with the Tel Aviv University, Beverly and Raymond Sackler School of Physics and Astronomy, Tel Aviv, Israel (telephone: 03-640-8303, e-mail: yiftahsi@post.tau.ac.il)

R. Ball, J. W. Chapman, T. Dai, C. Ferretti, D. S. Levin, C. Weaverdyck and B. Zhou are with the University of Michigan, Department of Physics, Ann Arbor, Michigan, USA.

J. R. Beene and R. L. Varner, Physics Division, Oak Ridge National Laboratory, Oak Ridge, Tennessee, USA.

P. S. Friedman is with Integrated Sensors, LLC, Toledo, Ohio, USA.

S. Ritt is with the Paul Scherrer Institute, Switzerland

hundred volts. A PDP in the simplest configuration i.e. matrix configuration, consists of two sets of parallel electrodes deposited on the surface of glass plates. The PDP is sealed with the two glass plates facing each other and their electrodes are orthogonal. The gap separating the two plates is filled with a Penning gas mixture, typically Xe, Ar and Ne. When a voltage pulse is applied between two electrodes a single pixel at the intersection of two perpendicular electrodes is illuminated. The voltage pulse leads to the breakdown of the gas and to the formation of a weakly ionized (only a small fraction of the atoms are ionized) plasma which emits visible and UV light. One way to way to construct a PPS cell is a coplanar arrangement of the electrodes. In this configuration, the plasma is struck between two parallel electrodes on the front plane. Addressing is provided by electrodes on the opposite plate, which are orthogonal to the coplanar electrodes [5].

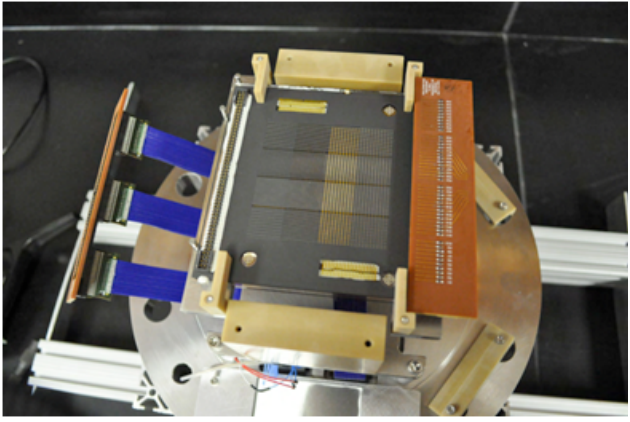

Fig. 3. Picture of the prototype PPS active area (44 cm$^2$) showing 16 different sectors (i.e. pixel geometries), instrumented on the motorized stage.

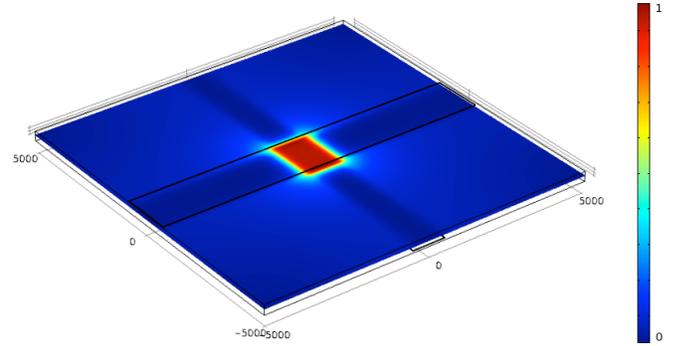

Fig. 4. *COMSOL* simulation of the electric field strength (normalized to one) inside one pixel in the commercial PDP.

We intend to utilize the structure of a PDP, but reverse the order of processes, i.e., instead of applying a voltage to produce light emission via a plasma discharge, we will arrange for ionization by radiation entering a PPS cell to cause a plasma discharge that we will detect electrically.

### A. Prototype PPS test chamber

In order to survey the general parameters of detector geometry, materials and gas mixture, we are staging a broadly configurable PPS test chamber. This test chamber is a vacuum vessel with ports for gas supply, gas exhaust, and electrode feedthroughs. Fig. 2 and Fig. 3 shows the prototype PPS active area, a 44 cm$^2$ platform, stepper motor translatable along the vertical axis toward or away from a glass electrode window, which serves as the stage upon which PPS test cells are mounted. Initial PPS cells employ low-cost glass substrates upon which are laid down by screen printing and photo-lithography techniques the surface-discharge electrodes. A screen printing process can also be used to produce the embedded resistors. These resistors are needed to current limit and bleed the pixel discharge. The test chamber is connected to a dedicated gas mixing system which can integrate up to four gas components in high precision. First measurements with the prototype PPS test chamber (in the University of Michigan in Ann-Arbor) started in October 2011.

### B. Simulations

Our simulation effort includes (both for the prototype PPS and the commercial PDPs)

*1) COMSOL[8]:* electric field and charge motion inside the pixels, electronic properties of the different components (e.g. capacitances and inductances of the pixels). Fig. 4 shows that the electric field is confined to the pixel area, this implies that the only active area in the panel is the area of the pixels.

*2) SPICE[9]:* Simulations of the electrical characteristics of the signal induced in the panel during discharge. Fig. 5 shows the schematic of one cell in the panel, this includes all the stray capacitances and inductances, lines resistance etc. The parameters in the *SPICE* models were determined with our *COMSOL* electrostatic model. The full *SPICE* model connects all the neighboring cells to form a large array of pixels.

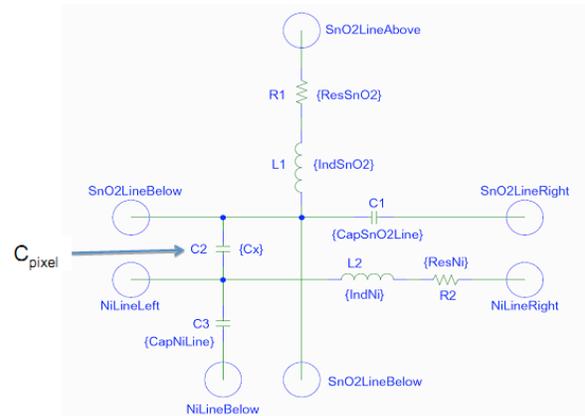

Fig. 5. *SPICE* model of one cell in the commercial PDP, the parameters C, L and R are capacitances, inductances and resistances of a single cell coupled to it's neighbors. Circles represent the cell's connections to other parts in the panel.

Fig. 6 shows the signal induced in one pixel while undergoing a discharge and Fig. 7 shows the signal induced in the neighboring pixels. The signal induced in the neighboring pixels is due to the capacitive coupling between all the pixels in the panel and not due to any discharge spreading. Since the signal induced in the neighboring pixels is positive in comparison to the negative signal induced in the discharging cell, the determination of the location of the discharging pixel should be straight forward (and is).

*3) GEANT[10]:* The active area in the panel is the gas volume between the orthogonal electrodes, which is covered by a 3 mm thick glass plate. In order to determine the response to radiation we simulate the energy loss and scatterings occurring prior the entrance to the panel. Fig. 8 shows the energy spectrum of betas entering the active area of the pixel, originally emitted by a $^{90}$Sr source.

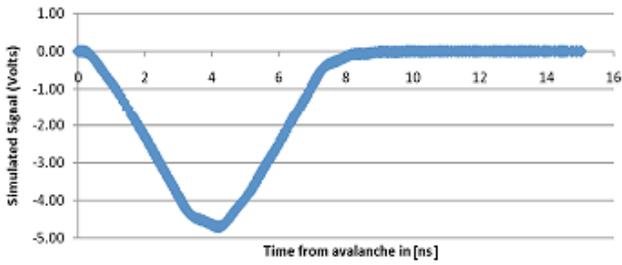

Fig. 6. SPICE simulation result for the output pulse from a discharging cell.

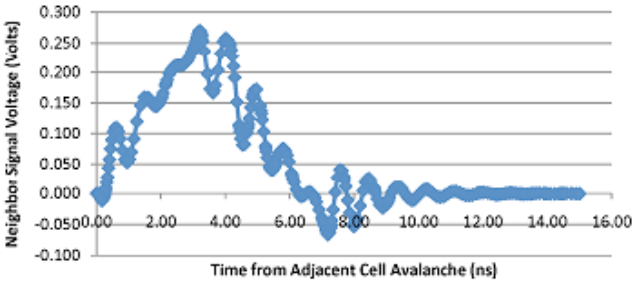

Fig. 7. SPICE simulation result for the signal induced in the neighboring pixels.

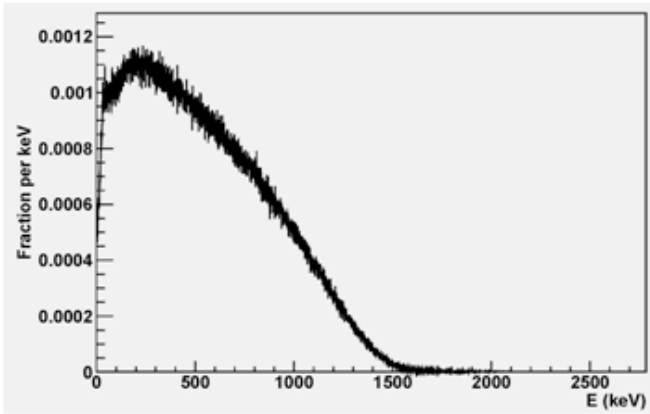

Fig. 8. Simulated spectrum of the radiation entering the panel from a $^{90}$Sr source.

## II. PDP LABORATORY EFFORT

As a preliminary step we study the behavior of generic matrix DC-PDPs. These panels are a monochromatic (i.e. no phosphor coating inside the panel cells) simplified version of the common matrix configuration PDP, of which, one is shown in Fig. 9 . There is no dielectric barrier caging every pixel, these panels are simply a matrix of anodes and cathodes with a roughly 300 micrometers gap filled with gas in between. Indeed, these modified commercial PDPs do produce signals when exposed to a radioactive source or when being traversed by a cosmic muon.

### A. Experimental setup

In order to explore the behavior of these PDPs under various kinds of radiation we have constructed two test benches (one in the University of Tel-Aviv and one in the University of Michigan). Each test bench includes a gas system which can refill the panels with premixed gases at a desirable pressure, mechanical support, a triggering system and DAQ.

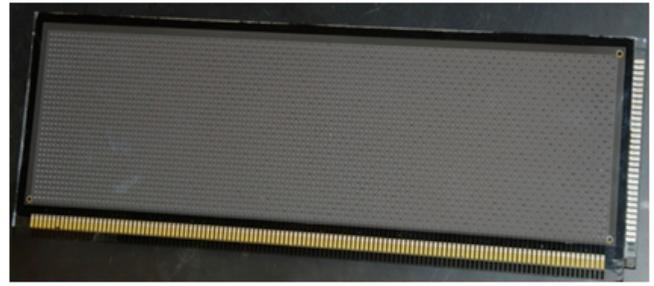

Fig. 9. A picture of a commercial PDP we are testing. Dimensions of this panel are 35.0cm×10.5cm.

*1) Triggering:* Triggering is being done with a set of small (few cm$^2$) scintillation pads and a dedicated hodoscope which includes two, thin scintillation pads and is designed especially for beta emission from a $^{90}$Sr source.

*2) DAQ:* Our laboratory effort includes the characterization of the signal induced in the panel during discharge. In order to do it we are using two, five GHz digitizer based on PSI's DRS4 chip [7]. For measurements of the rate of discharges we are using both the digitizer mentioned and a set of discriminators and counters. With the two digitizers (four channels each) it is possible to read a four by four array of pixels simultaneously, thus achieving a 2D position measurement of radiation traversing the panel. When moving to bigger arrays of pixels we use logic units connected to the discriminators triggered by the described set of scintillation pads. Fig. 10 shows one set up we are using in cosmic ray muons measurements.

*3) Gases:* As a part of our research we are investigating the panel response to radiation with various gasses in different pressures. We found that the panel is highly sensitive to the purity of the gas mixture, in order to prevent the contamination of the gas inside the panel we had constructed each of the gas systems solely on ultra-high purity stainless steel and other low outgasing materials, furthermore, a baking procedure for the panel is used when we change the gas content of the panel. Currently we are exploring the behavior of different gases mainly, Ar+$CO_2$ (at concentration of 93% - 7%), Ar+$CF_4$ (at concentration of 99% - 1%), Ar, $CF_4$, $SF_6$ and Xe.

### B. Response to radioactive sources

We have found that the panel responds to the radiation emitted from $^{90}$Sr and $^{106}$Ru sources, with all of the tested gasses and in pressures ranging from as low as 100 Torr to slightly below room pressure (The tested PDP is designed to work in low vacuum and will break under positive pressure). Fig. 11 shows a signal induced in the panel by $^{90}$Sr source placed above the panel. The gas content is Xe at 600 Torr and the operating voltage is 1210 volts. The signals we observe from all the tested gasses have large amplitude (at least few Volts) i.e. no need for amplification electronics. For each gas the shape of the induced signals are uniform. The leading edge rise time is few ns (at most). The observed pulses associated to single pixels (the area of each pixel is about 1 mm$^2$) with minimal discharge spreading between pixels (this was measured to be $O(2\%)$). $^{90}$Sr undergoes $\beta^-$ decay with decay energy of 0.546 MeV distributed to an electron, an antineutrino, and $^{90}$Y, which in turn undergoes $\beta^-$ decay with half-life of 64 hours and decay energy of 2.28 MeV, both $^{90}$Sr and $^{90}$Y are almost a perfectly pure beta sources. Due to this energy spectrum, it is highly unlikely (and practically impossible) for an electron emitted from this source to both trigger (pass through at least two scintillation pads) and enter the panel active area (pass the 3 mm glass layer). In principle, electrons emitted from $^{106}$Ru source have an energy spectrum reaching 3.54 MeV [11] which allows them to both trigger and induce discharges in the panel (electrons with this energy can penetrate the hodoscope and the 3 mm glass layer and ionize the gas inside the cell). Measurements of triggered $^{106}$Ru source are underway.

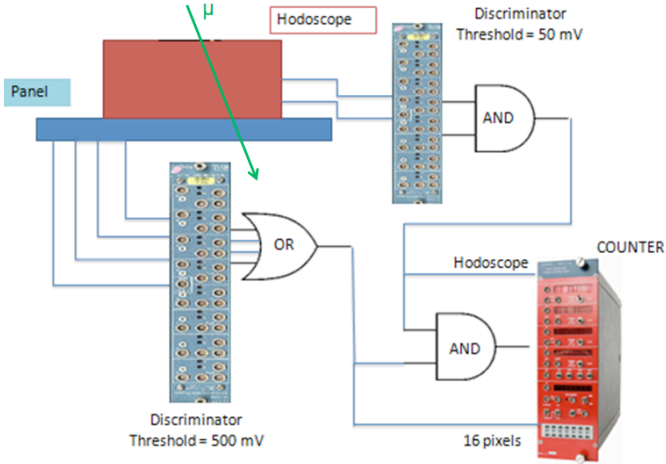

Fig. 10. Visualization of cosmic ray muon measurement setup, counting the number of signals induced in the panel which coincide with triggers from the hodoscope (associated with cosmic ray muons).

### C. Cosmic ray muons detection

Cosmic ray muons allow us to test the panel response to minimally ionizing particles. Using the setup shown in Fig. 10 we are able to associate signals induced in the panel with cosmic muons. With CF$_4$ gas at 600 Torr we have measured the panel total efficiency to be $O(10\%)$ for a voltage range of more than 50 Volts. The total efficiency is defined as the ratio of signals in the panel that coincide with the trigger and the total number of triggers (where all the triggers from the hodoscope are associated with cosmic ray muons). When taking into account that only the pixel area itself is active it yields that per pixel the efficiency to detect muons (with CF$_4$ gas at 600 Torr) is much higher $O(80\% - 90\%)$. We have also measured the elapsed time between the trigger (The time the muon passed through the panel in which Arrival time = 0) and the time of the signal in the panel. Fig. 12 shows the distribution of arrival times of 197 signals (i.e. cosmic muons). From this distribution and the fact that it's width is roughly 5ns as well as the fact that the leading edge rise time is $O(1ns)$, we can conclude that the timing resolution of this panel (in these operating conditions) is about 5ns.

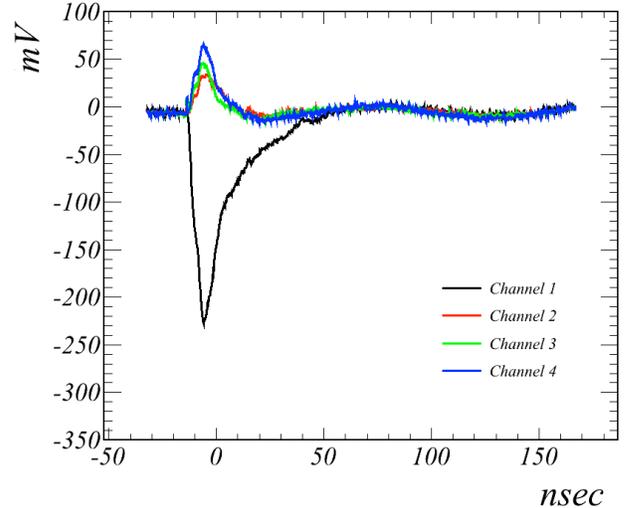

Fig. 11. A representative signal induced in the panel and attenuated to 0.01 of its value. Channel 1 (black) shows the discharge pulse (negative) while the other channels (adjacent pixels) show positive signals induced due to the capacitive coupling of the pixels in the panel.

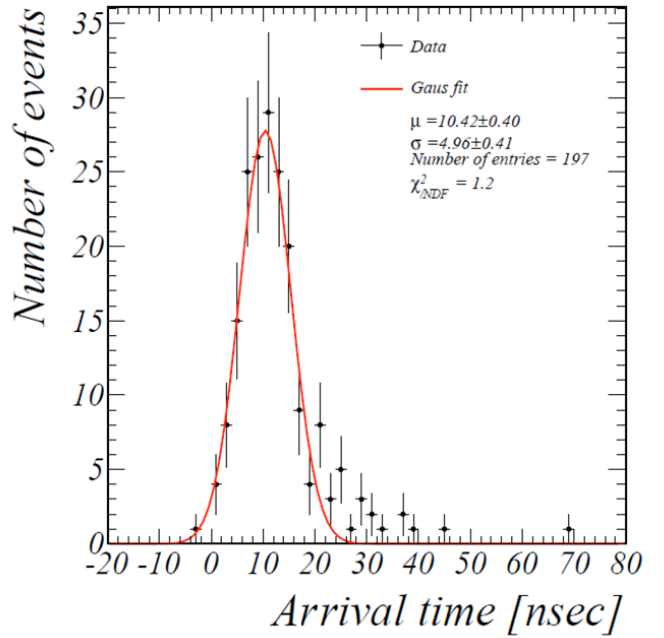

Fig. 12. Cosmic ray muons arrival time distribution. Gas content: SF$_6$ at 200 Torr and 1530 volts.

### III. SUMMARY

We have reported on the advances in the PPS development program, the coming along of a working PPS prototype, as well as first results of muon detection with a commercial PDP filled with different gases (mainly CF$_4$, SF$_6$ and Ar based mixtures). We have measured the panel's response to radioactive sources. We have also systematically started to characterize the pulses induced in the panel, thus expanding on our previously reported laboratory results regarding radiation detection with commercial PDPs [6].

ACKNOWLEDGMENT

The authors would like to thank Sergej Schuwalow at DESY for the hodoscope.